# The Early Years of Condensed Matter Physics at Illinois
## --in Celebration of the 80th Birth Year of
# Charles P. Slichter--

Leo P. Kadanoff
University of Chicago

## Charlie Slichter & the gang at Urbana

The 1950s-- and perhaps also the 1960s-- were very special times for the development of solid-state/condensed-matter physics. The University of Illinois at Urbana was at the center of these activities. In areas like NMR and superconductivity, methods were developed which would form the basis for the next half century of science and technology. Experimentalists, including Charlie and John Wheatley, worked hand in hand with theorists, including the incomparable John Bardeen. They worked cooperatively to develop ideas, often born in Urbana, but with godparents at Harvard and Moscow and Paris.

A characteristic style of broad collaboration and spirited exchange developed and spread from Illinois. This development was not an accident but the result of the vision of leaders like Wheeler Loomis, Fred Seitz, and later Gerald Almy[1]. The strong leadership saved the other scientists from expending their time on departmental decision-making. The style of the scientific activity was set by Fred, who strongly encouraged joint activities-- especially the interaction between between theory and experiment and between physics and engineering.   Fred encouraged comments from everyone, and helped everyone grow fast.

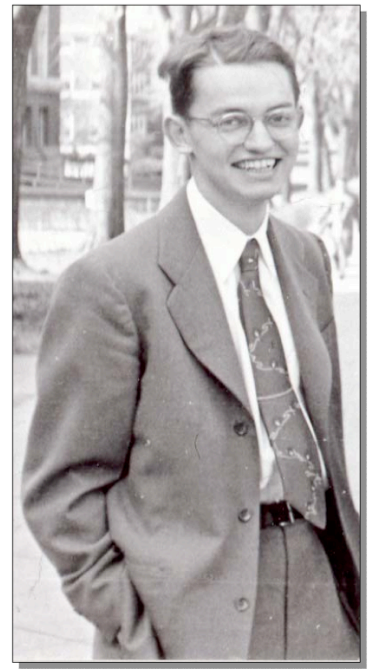

Charlie ca 1949

---

[1] For the Urbana audience, I put Urbana people in red.

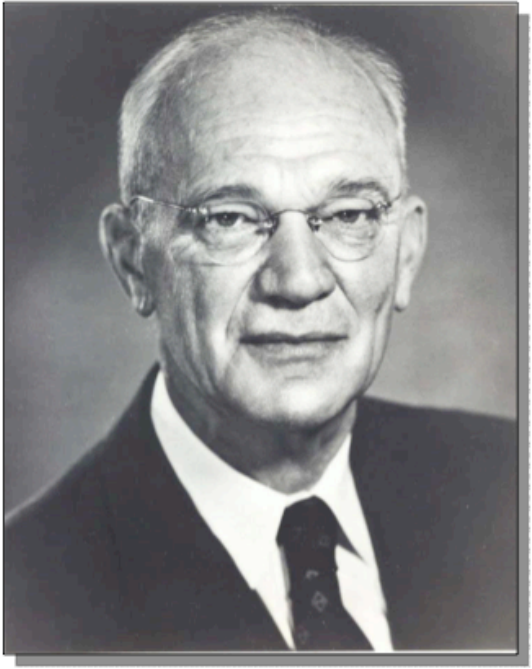
F. Wheeler Loomis
1929–1957

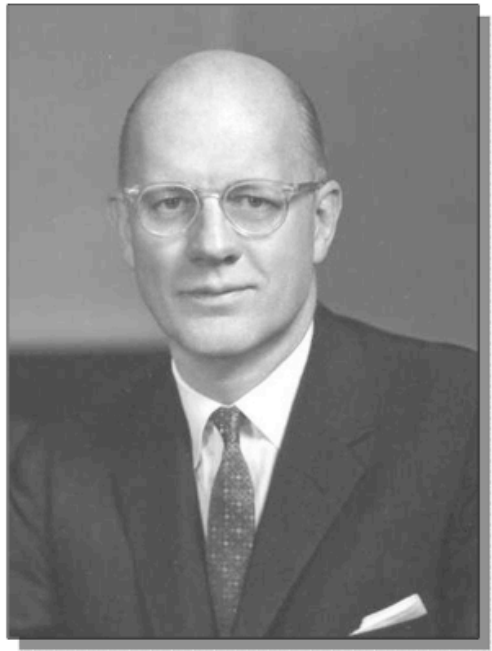
Frederick Seitz
1957–1964

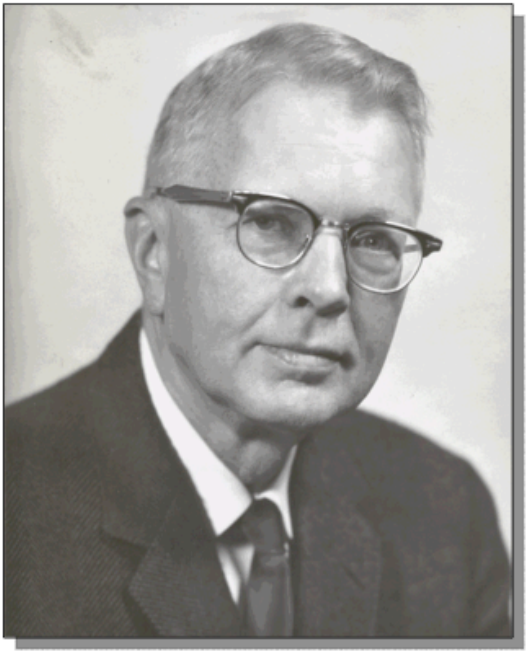
Gerald M. Almy
1964–1970

Department Heads, Physics Department, Urbana

# Purposes

This talk is intended to celebrate the times just after the arrival of Slichter and Bardeen to Urbana, the ideas which arose in that era, and especially the young people who--like Charlie and his students-- grew up then.

...... The NSF, in its wisdom, requires all of us who work with their money to stick to a two-fold set of purposes. Roughly speaking, our work has to have intellectual merit and also be good for somebody.

We can all feel sure that the work of the department in the era I shall describe, and most particularly the work of the two mythic figures, Charlie and John Bardeen, met these criteria. More important, these teachers set a high standard of decency, truth, and ethical conduct which we might wish to see more emulated in our present era.

## Our story starts with ancestors in Madison, Wisconsin.

Charlie is the grandson of Charles S. Slichter. During his 48 years at the University of Wisconsin, Charles became successively assistant Professor of Mathematics, Professor of Applied Mathematics and finally in 1921, Dean of the Graduate School. He wrote

I started off with four daily classes of forty freshmen each and an advanced class of two students, which soon dwindled to one ... I was so unsophisticated that I could think of no finer job than to teach mathematics to freshmen. Until quite recently I thought that everybody was of the same opinion. It was a shock to me to learn that some think that there is a higher job[2].

Charles' father, Sumner H. Slichter, was Lamont University Professor of Economics at Harvard. And his uncle, Louis B. Slichter, was a very distinguished geophysicist at UCLA. So the

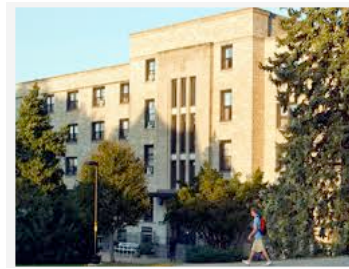

Slichter Hall, University of Wisconsin

---

[2] Charles S. Slichter, *Science in a Tavern*, University of Wisconsin Press, Madison, pages 177-178 (1966)

Slichters are part of American intellectual aristocracy.

The Bardeens are something special too...Charles Russell Bardeen, John's father, was the first person to graduate from the medical school of Johns Hopkins University. He then created a new medical school at, where else, Wisconsin and became its dean. John's mother Althea Harmer was an educator, who worked for John Dewey until Dewey had a falling out with the University of Chicago. John's grandfather, Charles William Bardeen, was also an educator and a publicist for better education. He wrote a book on rhetoric, and also *Little Massachussetts Fifer*, dairy entries and memoir of a roguish 14 year-old. (John Bardeen was, as I knew him, as far from roughish as one can get.)

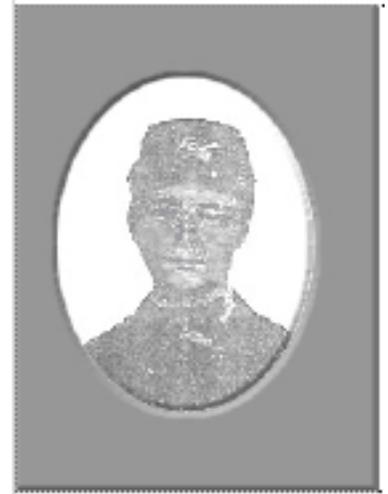

Charles William Bardeen

## Back to Illinois

I discreetly divert my eyes from Illinois and Illinois physics until the year 1929 when Wheeler Loomis is brought in to make the Physics Department at Illinois into national quality and prominence. The experience of the chemistry department under the direction of my old neighbor Roger Adams suggests that it can be done. But depression and shortage of funds impede Loomis' work until 1937. The war interrupts further development of the department, with Loomis himself going off to help develop radar at MIT's rad lab, and his Urbana team being mostly scattered.

In 1948-49 Loomis and his newly recruited graduate dean, Louis J. Ridenour, pulled off "a major feat: by a suitably arranged 'package deal,' he attracted Frederick Seitz, then Head of the Physics Department at Carnagie Institute of Technology, and his associate Robert Maurer, plus" the young instructors David Lazarus and Dillon Mapother, and in parallel brought Charles P Slichter from Harvard. The next three years brought to the faculty James Koehler, David Pines, Hans Frauenfelder, John Wheatley, and John Bardeen. These nuclear, atomic, and solid state physicists would, in the next years, build something new, "condensed matter physics".

## Back to Charlie (and Wisconsin)

CPS got all his degrees from Harvard. His undergraduate adviser was John Van Vleck, America's first quantum theorist. I return to the Wisconsin story for a moment: Professor Slichter's grandfather, in his first official act as Head of the Math Department at the University of Wisconsin, hired John Van Vleck's father, Edward, to be Professor of Mathematics in 1906. [Charlie] Slichter's parents became close friends with Abigail and John Van Vleck and lived a few blocks away from each other in Cambridge, often reminiscing about their Madison days. ... John Van Vleck served as [Charlie] Slichter's undergraduate adviser and encouraged him to continue his graduate studies at Harvard, suggesting he contact Edward Purcell about a thesis topic on electron spin resonance of paramagnetic atoms....

### Ms Celia Elliott adds:

I have a copy of a letter dated January 11, 1929, from Charles S. Slichter, Dean (U. Wisconsin Graduate School), to the "Senior Tutor" of Trinity College, Cambridge, forwarding "various papers relating to the application of Mr. John Bardeen for a research studentship in Trinity College ." Slichter adds: "I should be glad to have his application seriously considered."

### And also:

I have another longer letter from John Van Vleck to R.H. Fowler, recommending Bardeen for the Trinity scholarship. Van Vleck writes, "Mr. Bardeen is an exceptional student, unquestionably one of the two or three best I have ever had. He is taking my course in quantum mechanics, and grasps the subject so quickly that I feel that he is at times bored because I cover the ground so slowly, and is never forced really to exert himself in order to easily lead the class. .... I can recommend Mr. Bardeen without reservation as an unusually strong American candidate for the studentship."

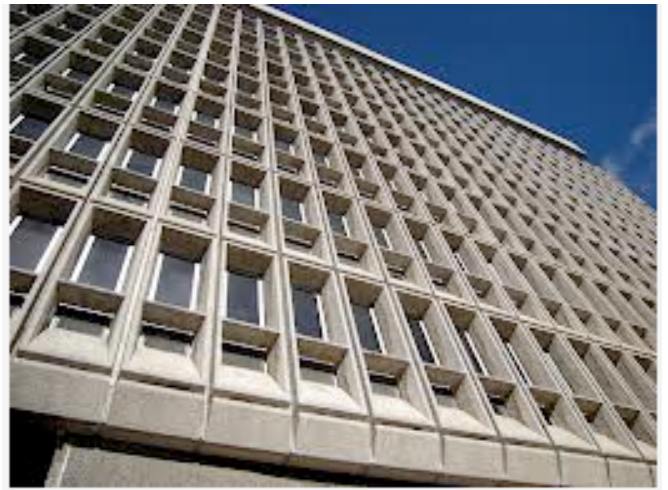

Van Vleck Hall
University of Wisconsin
named after John Van Vleck's father,
a Wisconsin mathemtaician

Scrawled on the bottom of the letter is "Dear [illegible], Will you add this to any dossier of Bardeen you have. Van Vleck I know well.  He is a man of sound judgment and European standards.  We may be fairly confident that this [illegible] student Bardeen is pretty good.  RHF"

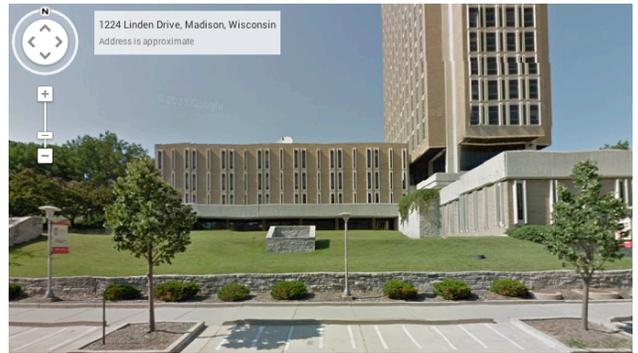

Bardeen Medical Laboratory, Madison Wisconsin

People who like closure will be happy to recall that Brian Josephson, another exceptional physicist, was hired as a research assistant professor by John Bardeen.  I recall his stay in Urbana quite fondly. Brian later became a fellow of Trinity, where he works to this day.

## Fast Forward;

Our own CPS arrived at Urbana, in 1949 fresh from a thesis at Harvard on electron spin resonance, ESR. At Harvard, he had worked together with a group  involved in both ESR and also in the related subject nuclear magnetic resonance, NMR. These acronyms describe techniques for studying what happens inside....  Inside what?  Inside anything: an atom, a liquid, a solid, or ---under the name magnetic resonance imaging (MRI)-- inside you or me. CPS's academic parents Edward M. Purcell and Norman Ramsey got these subjects going and CPS and his contemporaries George Pake, Nico Blumbergen, Bob Pound, and Herb Gutowsky were to play a large role in developing these and related areas.  Herb and Charlie were to lead the Illinois effort in this area.  Their interaction was enhanced by the equal importance of chemistry and physics in this area of work, and by Charlie's solid capabilities as a theorist that enabled him to fill in a gap in that day's chemistry department.

CPS was (and is) a first rate teacher
Gerald Almy (then department head) re Charlie:
"One measure of the academic scientist's contribution to the advancement of his field is quality and continuing productivity of the students whom he guides in research to the level of the Ph.D. degree. Professor Slichter has been singularly successful in developing his students into first-rate physicists."

Charlie's Students of that era
Dick Norberg "magic angle spinning"
Don Holcomb "confirmed Korringa relation"
Erwin Hahn "spin echos"
Meyer Bloom "outdid CPS in quadrapole resonance theory"
Chuck Hebel
Burton Muller
Bob Schumacher
Tom Carver

## The hat trick[3]

Looking back, one can see that in the period 1953-1956 CPS and his group produced three physics contributions of the first level of importance.
1. Al Overhauser, an Urbana postdoc, had done a thesis predicting a subtle but useful relation among different resonances. In the thesis and thereafter, Al made absolutely outrageous statements about physics, which subsequently proved right. "Proved right" is a passive construction. Actively speaking, CPS and his student Tom Carver did a series of experiments in sodium directed at Al's predictions. The result? CPS's words: "The big shots in magnetic resonance did not believe Overhauser's prediction, so it was a source of great excitement when we showed he was correct." Al's prediction, now called the "Overhauser effect" got me more than 800 hits on google. Charlie and Tom saw it first.

---

[3] The term hat trick comes from the English game of cricket and refers to a bowler who takes three wickets with three successive balls…[and was] awarded a new hat by his club as a mark of his success…. [This term was] first recorded in print in the 1870s, but has since been widened to apply to any sport in which the person competing carries off some feat three times in quick succession.

2. A little later, CPS and David Pines were talking about measuring a fundamental property of the electron gas, called the spin susceptibility. Nobody knew how to measure it, except one CPS who understood the theory of how a material will respond to an external stimulus.
So CPS and his student Bob Schumacher did the measurement of the spin susceptibility. To carry out and interpret the experiment, Charlie needed a piece of the rapidly developing theory of "elementary excitations" Urbana, and especially John, David, and Charlie, were major actors in the development and interpretation of this theory, which was also beginning to emerge from the work of Landau (in Moscow) and others in Paris and New York. The experiment thus required for its construction and interpretation big chunks of theory. The highly interactive environment of Urbana was beginning to pay off.

3. At that time, Illinois was abuzz with work on superconductivity theory. Bardeen had just put together a piece of thinking that showed how superconductors might possibly have an energy gap in their spectrum. CPS soon recognized that nuclear magnetic resonance would measure nuclear spin-lattice relaxation time and that relaxation would show a clear signature of the gap. So before the work of Bardeen, Cooper, and Schrieffer, Charlie and his student Chuck Hebel set out to do the technically difficult experiment. They got a surprising result in which the relaxation rate increased by a factor of two as the temperature was lowered into the region below the critical temperature. This total surprise became and object of considerable joy as the BCS theory was later developed and showed basic agreement with the work of Hebel and Slichter (and later of Al. Redfield).

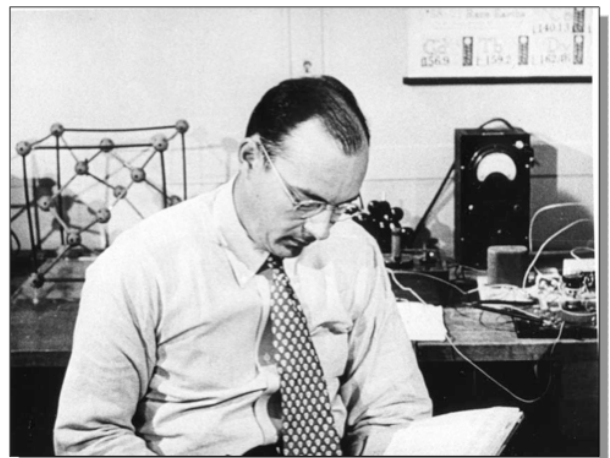

John Bardeen

In describing this third piece of the "hat track," CPS said:
Al Redfield, Chuck Hebel, and I look back on our experiments with very special pleasure. It was a time of great excitement. Explanation of superconductivity had eluded scientists for many years. To feel that one might be close to a solution was a great impetus for our experiments.
....the most remarkable thing was to be instructed in the theory at the very time that its authors {BCS} were developing their calculations. What remarkable openness! What remarkable kindness to take the time to help us when they were racing to explore the consequences of their discovery!
International Journal of Modern Physics B Vol. 24, Nos. 20 & 21 (2010) 3787–3813.

Back to the theorists: In the meantime, ideas and accomplishments were boiling out of the Illinois condensed matter group. The crucial idea of "elementary excitation" is that every solid contains rather simple stuff moving through it, for example electrons and sound waves. In this context, the sound wave is made up of bundles called "phonons". The next step, put together by Bardeen and Pines is how to describe what happens when a phonon bumps into an electron. The answer is that the phonon might perhaps be absorbed, and the process is described by specifying the chance for absorption, in terms of what is called the "effective interaction" between the electron and the phonon. Bardeen and Pines calculated the effective interaction and emphasized its importance. This interaction forms the basis of the theory of Bardeen Cooper and Schrieffer, which comes up next.

**Superconductivity Theory.** Bardeen, his postdoc Leon Cooper, and his graduate student Robert Schrieffer were working together on the correct description of superconductivity, a problem which remained unresolved for forty-five years. At very low temperatures some metals, including aluminum, have a very rapid change in their properties including an abrupt decline in their resistance to the flow of electrical current. John felt that this problem was truly worthy of his intellect and training.

As described in *True Genius*, John had left Bell somewhat dissatisfies, a dissatisfaction only whetted by his first Nobel prize. John wanted and needed an important new accomplishment. For this little group, John provided, in addition to leadership, many years of hard thought about all aspects of the problem. Leon figured out how the electrons hooked together to form what are now called "Cooper pairs", Bob put together the quantum description of the problem.   Together, in one of the great accomplishments of the last century,  they figured out how superconductivity arose and the effects it would have upon a metal.

Their theory of superconductivity, known by their initials as the BCS theory, did not approach the commercial importance of John's earlier work on the transistor. Nonetheless,  it was clearly the most important single scientific theoretical advance in the period after world war II.  It showed us how the matter around us could make use of quantum phenomena to do all sorts of wonderful things. Indeed the very words "condensed matter" that we use to describe our subdiscipline resonate in our ears because of their reminder that Cooper pairs  condense[4].  into a kind of soup, which then does the aforementioned wonderful things,

In this celebration we should of course remember that no one person, no three people, make a science. The BCS work was complemented by the important advances of Landau & Ginzburg and Abrikosov, of Bogoliubov, of Blatt, Butler, & Shafroth, and others.  But one piece of the action belongs to BCS and Urbana alone.

---

[4] Somebody might wish to write an essay about the words in our field: "stimulus", "response", "susceptibility", "excitation", "jamming", "frustration".... all technical words with specific meanings a little different from the popular ones.

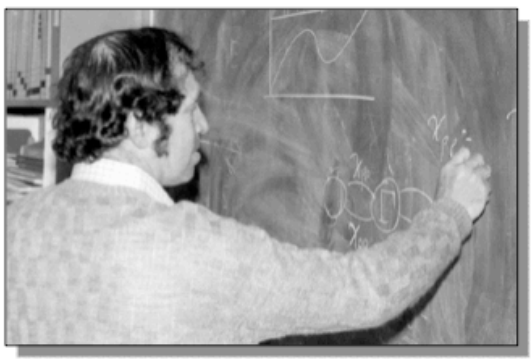
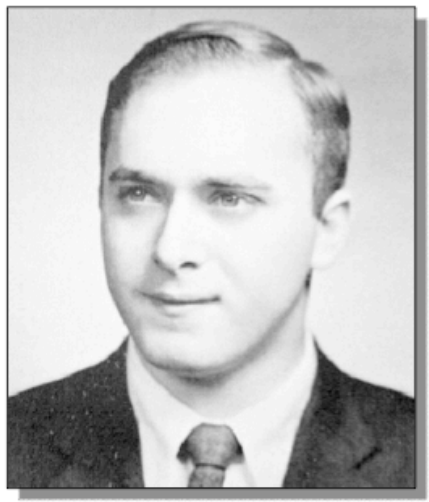
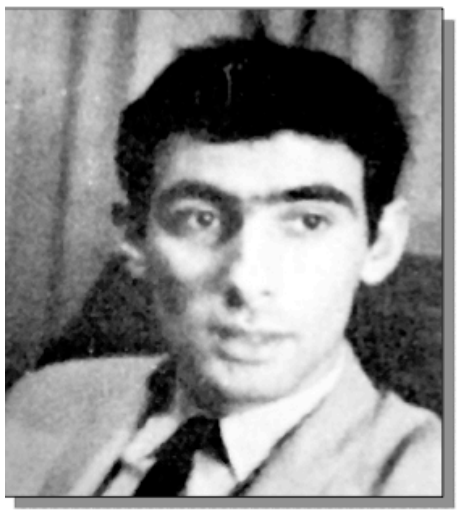
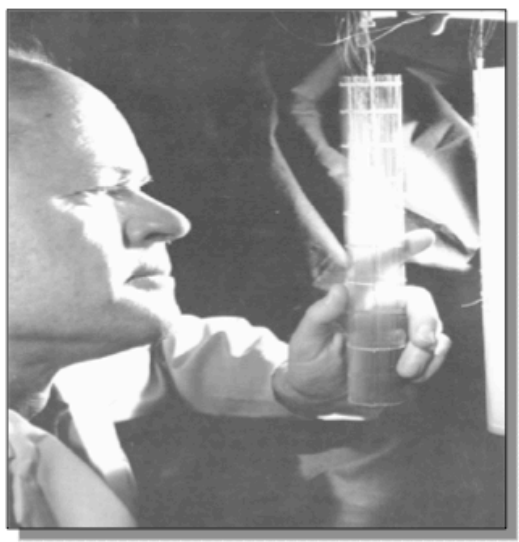

Pines, Baym, Kadanoff, Wheatley

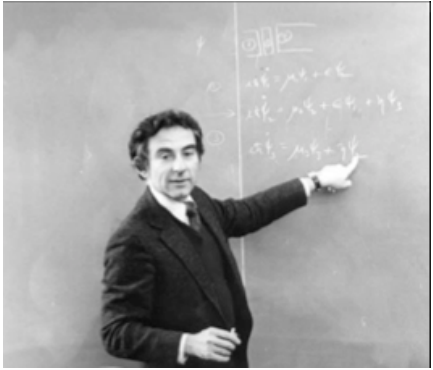 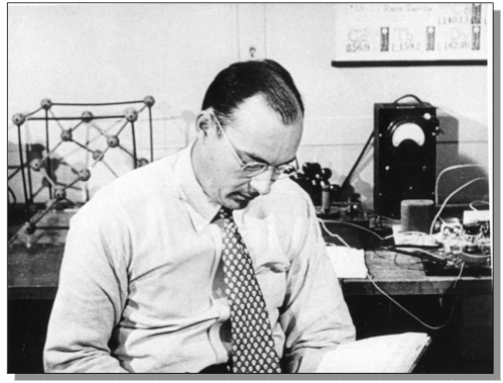 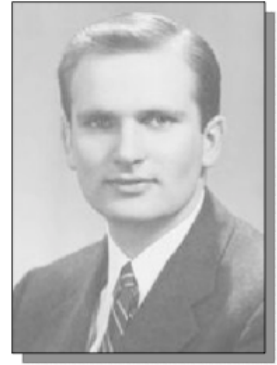

Leon Cooper    John Bardeen    J.Robert Schrieffer

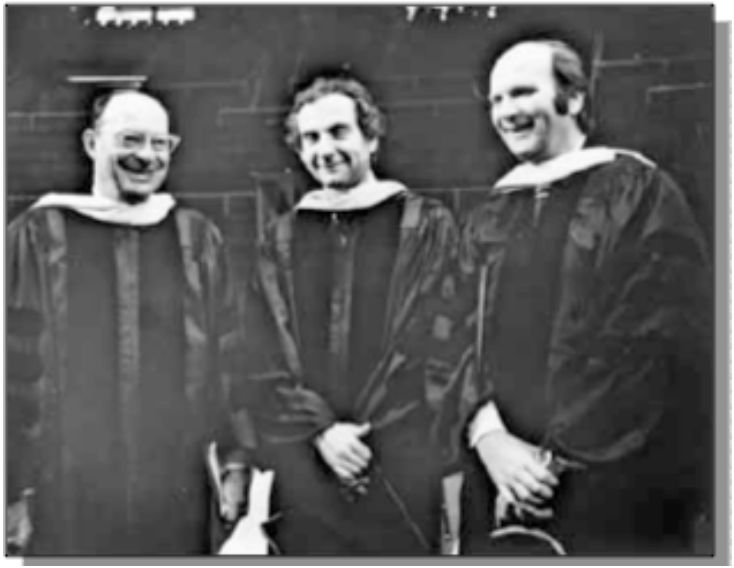

All together in Stockholm

## And More....

Urbana physics reached a high point in the superconductivity studies, and then continued on through several other very notable achievements. The two great ideas circulating through BCS and the other work of the period are: "elementary excitation" and "order parameter"(see below). John Wheatley in his work on Helium three and Bardeen, Baym, and Pines in their work on Helium three-four mixtures did especially elegant and important work on excitations. These works were respectively peaks of the experimental and theoretical arts.

My own Urbana work on phase transitions owes major debts to my Urbana environment. I built a theory on the notion that the system contained an "order parameters" and several different "elementary excitations." . The work was based upon the ideas of "effective interactions" which arose from the Bardeen-Pines study of the electron-phonon interaction.

"Elementary excitations" and "order parameters" also because the basis of the present day "standard model" which defines the field of particle physics.

## And More....

Very important early work on the behavior of the order parameter belongs to the Laudau school: Lev Landau, Vitaly Ginzburg, A. A. Abrikosov. The highest peak of the later studies was put together by Brian Josephson. Brian's brilliant ideas were first greeted by Bardeen's skepticism, and later by John's very public statement that indeed Brian Josephson was right. This act of grace was one of Bardeen's highest public moment.
Charlie's highest moments take place every day, in his very special style, treating the great and the little equally to his good humor and good fellowship.

One example involved an ugly argument which arose at the Ph.D. exam of one of my students, a Korean. A member of the committee refused to pass the thesis. When pressed by Charlie, he said the English was poor. Some sort of prejudice or irrationality seemed to be involved, perhaps toward the student, or me, or our ethnicities. Charlie first offered to arm-wrestle with the objector. After that partially facetious offer was refused, CPS suggested another, face-saving, resolution: changing some wording in the thesis.  The meeting broke up without wrestling.

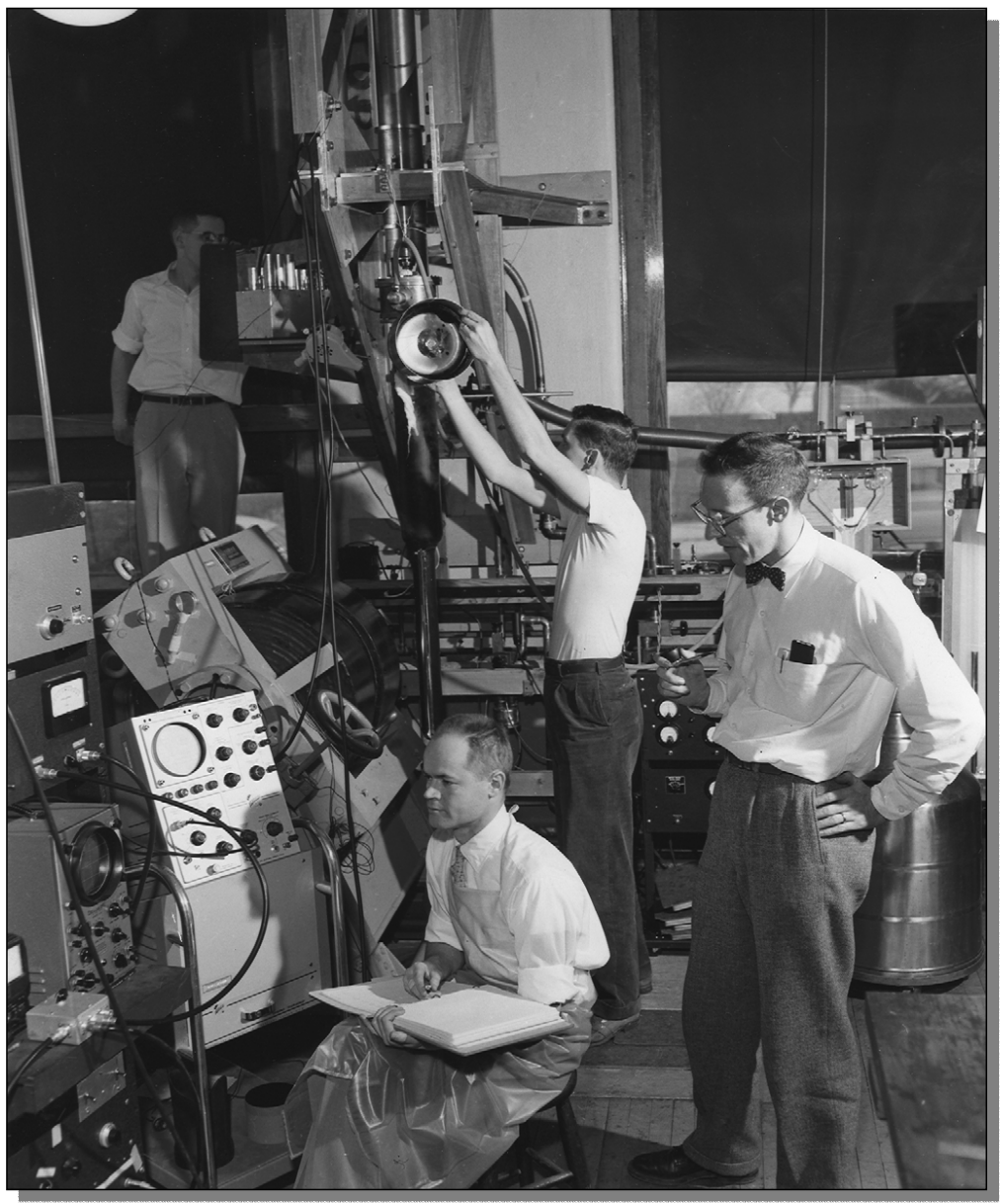

**Howard Hart, John Wheatley (seated), Thomas Estle, Dillon Mapother, 1957**

# A Wider Perspective

The department has a history of constructive engagement in national affairs.  Examples:

1951: Wheeler Loomis and Frederick Seitz, along with other members of the department, sign an open letter to President Truman, protesting the zeal with which hydrogen bomb research is being undertaken and urging caution in its development and use.

1954: Thirty-eight faculty members sign an open letter to the Atomic Energy Commission in support of beleaguered physicist J. Robert Oppenheimer and warn that "scientists of ability and integrity" will hesitate to accept advisory positions in the government if they were to be punished for voicing unpopular views.

John and Charlie were members of the Presidents' Scientific Advisory Committee, which provided advice to Presidents from Truman though Nixon.

As part of Charlie's public service role he served on the Harvard Corporation, Harvard's highest governing board, for twenty-five years. John helped found Xerox (called then, Haloid), and served on its board from 1952-1974.

# The Broader Stage--Continued

Given that thoughtful and decent people including CPS, Bardeen, Seitz, and Loomis were important actors on the national scene, and given that physics was in a position of unprecedented power, prestige, and wealth, it is perhaps tempting to look back at the 50s as "good times".  But, major problems certainly existed then.  The price of intellectuals' prosperity was a very close tie to the military and CIA.  Our aborted invasion of Cuba and our march up and back to Korea's border with China both served as reminders that governmental actions were not always right or wise. Physics' ties to the military might turn out to be a mixed blessing.

On the home front, Joe McCarthy, Senator from Wisconsin, reminded us that clear thinking was not always desired in Washington.

The present narrative totally lacks mention of women or blacks in Urbana physics. This lack accurately reflected the scientific demographics of the period.  In this period, prejudice was still openly practiced in academic hiring. But let's return to better things.

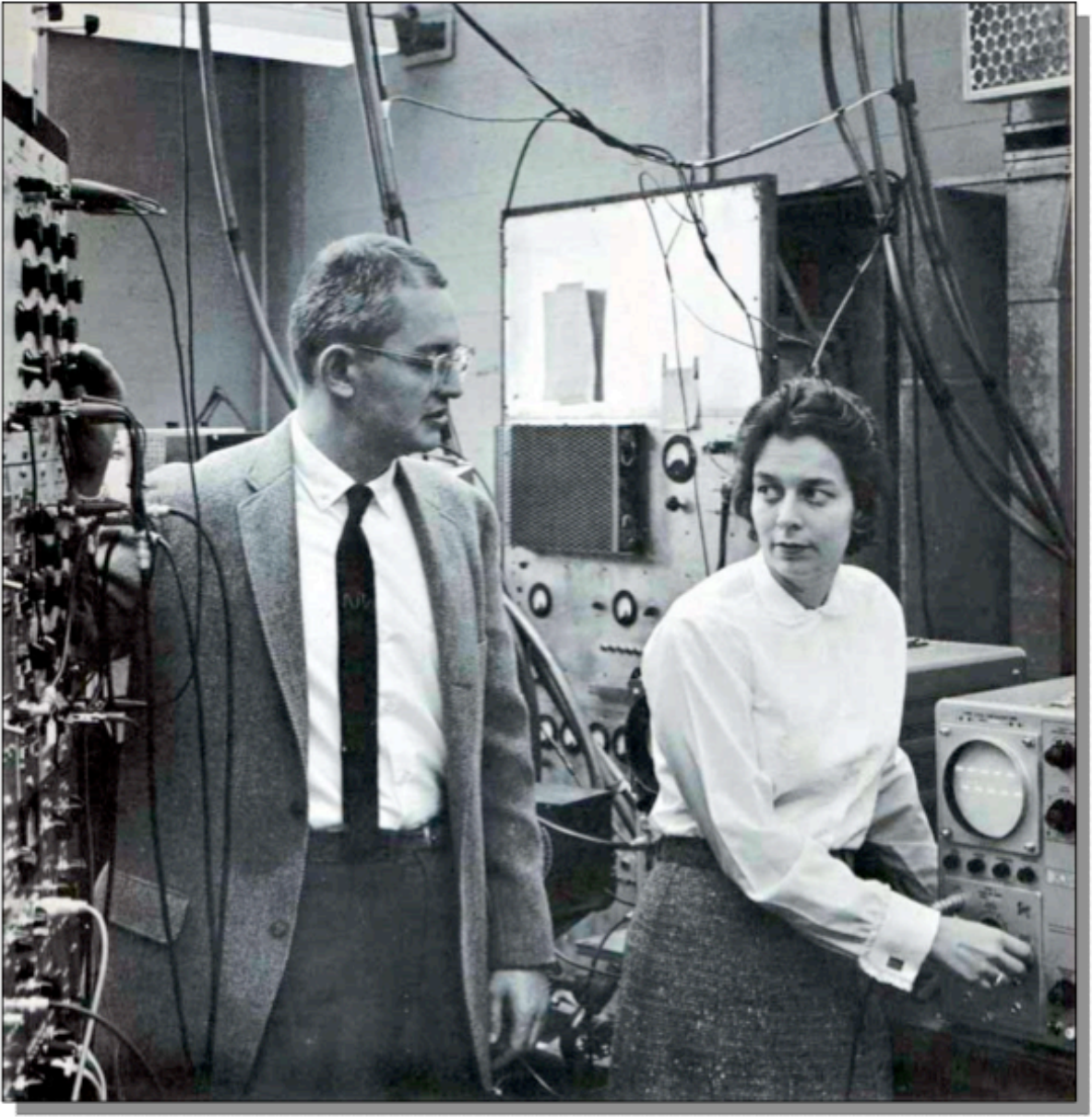

**CPS and Judy Franz, ca. 1964**

Judy got her PhD and went on to become executive secretary of the American Physical Society, a position from which she provided a very considerable leadership of the physics community. Charlie and John provided a tolerance and breadth of respect that was uncommon in the universities of the 1950s.

## John and Charlie provided lots of leadership

Example: I wrote a paper entitled "Failure of Electronic Quasi-particle picture for.....". This title was somewhat misleading because my paper only showed that the simplest version of " the electronic quasi-particle picture" was false. I knew then that the more sophisticated version remained correct. Charlie bawled me out for being misleading, saying that I was guilty of puffing up this work beyond its true worth and such puffery has no place in science. Unfortunately now it does have a big place in *Science*, and *Nature*, and *Physical Review Letters*, ....

Example: I brought a problem to John. I had a student who finished a piece of theory. As he was writing it up, we found that theory given in two preprints. What to do? Ignore the earlier work to get my student a Ph.D.? "No" was John's reply. "You can't do that."9 But we were at Urbana. We rewrote the paper to compare the other guys' theory to all the relevant experiments. The student got the Ph.D...and lots of citations.

## Conclusion

Our world today needs John Bardeen's statement: "you cannot do that". The people who put together Enron's energy price manipulations should have heard that. Our leaders, take your pick, on the highest level should have heard that early and often. Schön and Batlogg should have been told "you cannot do that" long before they came to places of prominence in our scientific world.

Charlie was and is right about puffery. Misleading statements are wrong. The highest standards of truth should apply to our papers, our proposals, our statements about the value of what we do. When we employ distortions of the truth to support our own work or to raise money for our science, then we are violating the trust which must be the basis of science and of our field's relation to the public. Instead we scientists should try to set a good example for the society at large. On a larger stage, when distortions of truth are used to win support for governmental policies this is another violation of trust, and this violation weakens representative

page 18

government.   We have seen in our time the weakening of science, science's support, and our republic.

Paradoxically, while the 50s had very substantial unresolved and partially unexamined problems, it also had in many of its actors (like John and Charlie) the highest forms of idealism. This idealism included a very substantial effort to teach and apply high ethical standards. We celebrate the 50s (and the 60s), in part because these decades represent something now lost.  We need to learn to say again "you cannot do that."